# Room temperature multiferroicity in a transition metal dichalcogenide

G. Cardenas-Chirivi[1,2], K. Vega-Bustos[1], H. Rojas-Páez[1], D. Silvera-Vega[1], J. Pazos[1], O. Herrera[2], M. A. Macías[3], C. Espejo[4], W. López-Pérez[4], J. A. Galvis[5] and P. Giraldo-Gallo[1]

[1] Department of Physics, Universidad de Los Andes, Bogotá 111711, Colombia

[2] Faculty of Engineering and Basic Sciences, Universidad Central, Bogotá, Colombia

[3] Department of Chemistry, Universidad de Los Andes, Bogotá 111711, Colombia

[4] Department of Physics and Geosciences, Universidad del Norte, Barranquilla, Colombia

[5] School of Engineering, Science and Technology, Universidad del Rosario, Bogotá 111711, Colombia

(Dated: February 3, 2023)

The coexistence of multiple ferroic orders, i.e. multiferroicity, is a scarce property to be found in materials. Historically, this state has been found mainly in 3-dimensional complex oxides, but so far this state has still been elusive for the most widely studied and characterized family of 2-dimensional compounds, the transition metal dichalcogenides. In this study we report the experimental realization of multiferroic states in this family of materials, at room temperature, in bulk single crystals of Te-doped $WSe_2$. We observe the coexistence of ferromagnetism and ferroelectricity, evidenced in the presence of magnetization and piezoresponse force microscopy hysteresis loops. These findings open the possibility of widening the use and study of van der Waals-based multifunctional devices for new nanoelectronics and spintronics applications.

## 1. Introduction

Multiferroics are materials that exhibit different coexisting ferroic orders such as ferroelectricity, ferromagnetism, or ferroelasticity. Due to the coupling among the different degrees of freedom leading to these ordered states, the order parameters of one state can be controlled by tuning parameters different from their conjugate variable. For example, magnetoelectric coupling allows the control of magnetization via an electric field, or conversely, the control of electric polarization via magnetic field, for the case of ferromagnetic/ferroelectric heterostructures and multiferroics. Multiple applications in spintronics, data storage, actuators, among others (*1–4*) have been envisaged and already realized. The search for new and better multiferroic materials is arduous: ferroelectrics tend to be insulators as they need to preserve electric polarization (free charges in metals screen this effect) while ferromagnets are metals in their majority (*5*). In this sense, a guiding strategy that has been used to attain profitable multiferroicity is chemical doping of either a ferroelectric host or conversely, a ferromagnetic host. For example, replacement of the transition-metal ion in ferroelectric $LiNbO_3$ or $ZnO$ leads to magnetic order. On the other hand, chemical doping in complex oxides multiferroics has been effective in optimizing their properties (*6, 7*).

Historically, these phenomena have been found mainly in 3-dimensional complex oxides and perovskites, such as $Cr_2O_3$, $YMnO_3$, $BiFeO_3$, among others, or in heterostructures of ferroelectric/ferromagnetic thin films (*8–10*). The 3D character of the crystal structure of these compounds poses challenges in their incorporation into nanoscale multiferroic devices. Their nanostructuration requires advanced fabrication techniques, as well as induces physical effects inherent to quantum confinement that typically weakens the multiferroic states (*11*). In this sense, the use of multiferroics with an intrinsic 2D-crystalline structure is highly desirable for obtaining affordable and robust multiferroic nanodevices.

Recent advancements in the realization and study of intrinsically 2D-multiferroics have been done (*12, 13*), for example in the transition metal halide compounds such as $NiI_2$ (*14*) and $CrI_3$ (*15*), and in doped $In_2Se_3$ (*16*). However, multiferroicity has been elusive, until now, for the most widely studied and characterized family of 2D compounds, the transition metal dichalcogenides (TMD's), in spite of numerous theoretical works in this respect (*17, 18*). In recent years, for this family of materials, individual ferroic orders have been experimentally observed at room temperature (*19–24*). Ferromagnetism has been attained in $MoS_2$ and $WS_2$ due to defects (*21, 23*), and in $WSe_2$ via chemical doping in the transition metal site (*24*), among other examples. Ferroelectricity has been reported in bulk-$WTe_2$ (*20*), while for an odd number of layers, $WSe_2$ and $MoS_2$ are reported to be piezoelectric (*25, 26*). Despite the presence of these individual ferroic orders, there is no experimental evidence for simultaneous ferromagnetism and ferroelectricity in TMDs to date.

In this study we report the experimental realization of multiferroic states in TMDs, specifically in Te-doped $WSe_2$. The multiferroic state is observed at room temperature and in *bulk* single crystals, for Te doping above 7%. The multiferroic states are revealed in the observation of ferromagnetic hysteresis loops in the magnetization, and ferroelectric hysteresis loops in the phase and amplitude of piezoresponse force microscopy (PFM) measurements. In addition, the highest Te-doping samples studied (15%) show very large effective piezoelectric coefficients, comparable to the commercially used 3D-piezoelectric $LiNbO_3$ (*27*), and one order of magnitude higher than in $WSe_2$ and other pure TMD's nanolayers (*25, 26, 28*). In addition, we present density functional theory (DFT) calculations which provide insights about the possible mechanisms leading to the multiferroic state. These findings open the possibility of widening the use and study of van der Waals (vdW)-based multifunctional devices, and provide an affordable platform for the study of fundamental physics of ferroic orders on low-dimensional compounds.

## 2. Results

### *Raman spectroscopy and X-rays techniques*

Figure 1(a) shows a picture of representative single crystals of Te-doped $WSe_2$, with typical lateral sizes around 1mm, and a hexagonal morphology. A level of chalcogen vacancies is naturally developed during the crystal growth process, and crystals with a general stoichiometry $W(Se_{1-x}Te_x)_{2(1-\delta)}$ are produced. x represents the fraction of Se atoms that were replaced by Te atoms, and $\delta$ is the fraction of chalcogen sites that are not occupied by neither a Se nor a Te. The values of x and $\delta$ for the different crystals studied were determined through X-ray fluorescence spectroscopy (see methods, and table S1 for values). Samples with four different Te-doping values, up to a maximum of x=15%, plus undoped samples (x=0) with three different levels of chalcogen (Se) vacancies were characterized in this study. Two different types of X-ray diffraction measurements (powder diffraction, presented in figs. 1(d-f) and fig. S2; and single crystal diffraction, presented in table S2 and crystallographic information file - CIF - available in supplementary files) reveal that the crystalline structure of all the pure and Te-doped compounds is the 2H-structure (space group $P6_3/mmc$, # 194) (*29*). The most intense diffraction peaks in fig. 1(d) come from the {0 0 1} family of planes, given that the *a-b* plane of the crystals tends to align parallel to the sample holder (see methods). Close inspection to the {0 0 1} diffraction peaks (fig. 1(e) for the (0 0 2) peak) reveals a continuous shift to the left, and therefore, an increase of interplanar *c*-lattice parameter with increasing Te-doping. Single crystal diffraction confirms the increase in *c*-lattice parameter and, in addition, reveals an increase of the in-plane *a*-lattice parameter with Te-doping. Table S2 shows a summary of *a* and *c* lattice constants found by X-ray diffraction.

The 2H-polytype is also confirmed by Raman spectroscopy. Figure 1(c) shows the Raman spectra for the different Te concentrations studied (x=0-15%), around the $E_{2g}$ and $A_{1g}$ modes for 2H-WSe$_2$. A continuous decrease in the Raman shifts of both modes as tellurium concentration increases is observed, which also reveals a 2H-structure with continuously increasing *a*- and *c*-lattice parameters. Raman peaks associated with the 1T$_d$-polytype, which for pure WTe$_2$ appear around 150 cm$^{-1}$ and 200 cm$^{-1}$, are not present, implying that secondary phase domains with this structure are not detected. Our results of the crystallographic structure align with a previous study in W(Se$_{1-x}$Te$_x$)$_2$ which reports a pure 2H-structure for Te compositions below 50% (*30*).

### *Magnetic response*

Magnetization measurements as a function of magnetic field at *room temperature* are shown in figures 2(a) and S3. These reveal two contributions to the magnetic signal: a diamagnetic component and a paramagnetic or ferromagnetic component, depending on the chalcogen vacancy level. The paramagnetic behavior has been previously reported by other groups in WSe$_2$ and other TMDs (*21, 22*), and the paramagnetic or ferromagnetic contribution have been theoretically predicted (*31*) and experimentally realized in this family of materials (*21–24, 32–34*). Measurements of samples with different times of exposure to ambient conditions suggests that the diamagnetic component is highly dependent on the level of oxidation of the samples. For instance, freshly synthesized samples showed no or minimum diamagnetic contribution, whereas the same samples after being stored in soft vacuum for a couple of days showed a marked diamagnetic contribution (fig. S4). The paramagnetic/ferromagnetic contribution appears to be more robust to sample oxidation

(figs. S4 and S5, and table S3), and therefore the diamagnetic contribution of all curves in fig. 2(a) was subtracted in order to focus on the paramagnetic/ferromagnetic component. Samples with no Te doping and low concentration of chalcogen vacancies (black curve in the inset to fig. 2(a)) show a paramagnetic behavior, with a saturating magnetization above fields of the order of 150 Oe. Te-free samples with an increased number of Se vacancies showed weak hysteresis loops with small coercive fields, indicating weak ferromagnetism (see red curve in the inset to fig. 2(a), and fig. S4(a-c) for other vacancy levels). For all the Te-doped samples studied, the ferromagnetic hysteresis loops became more notorious, reaching a maximum coercive field of 700 Oe for the sample with x=7.6% and δ=38%. Figures 2(b,c) plot the coercive field for the different samples studied, as a function of Te doping (x) and chalcogen vacancies (δ), respectively. The coercive field as a function of chalcogen vacancies grows monotonically, in contrast to its behavior as a function of Te doping. This suggests that magnetism is driven by the amount of chalcogen vacancies. Nevertheless, the apparent strengthening of the ferromagnetism in the Te-doped samples, in contrast to the undoped samples, suggests some role of the Te atoms. We argue that this is due to the fact that the presence of Te in the synthesis promotes the formation of chalcogen vacancies (Table S1). For instance, all the Te-free synthesis reached maximum vacancy levels of 12%, even though for some of them nominal vacancy values of 50% were aimed. Whereas for the Te-doped synthesis, measured vacancy levels above 12% could be easily reached, with a maximum obtained vacancy level of 38%.

### *Piezo- and ferro-electric response*

Piezoelectricity in these compounds was investigated using an atomic force microscope (AFM) through the Dual AC Resonance Tracking (DART) piezo force microscopy (PFM) technique (see methods). Freshly-cleaved samples were measured in an inert $N_2$ atmosphere in order to prevent surface oxidation and minimize electrostatic effects. This technique measures the amplitude of elongation for a given applied AC bias, and therefore it is a direct measurement of the piezoelectric response. Figures 3(a-c) show DART piezoresponse maps for different values of the AC bias voltage, in single crystals with three different doping levels. The color in each plot represents the amplitude of out-of-plane deformation, which spatial average increases with the AC applied bias. The spatially-averaged piezoresponse as a function of AC bias, presented in figure 3(d), is linear for all samples. The slope in each curve is equivalent to the effective piezoelectric constant $d_{33}$ (see methods, and section S4). The two lowest Te-concentration samples exhibit comparable values for their effective $d_{33}$, of the order of 4 pm/V. This value is similar to the piezoelectric constants previously reported for TMD monolayers (*26*, *35*), and higher than the $d_{11}$ coefficient reported for a $WSe_2$ monolayer (*25*, *28*). For the highest Te-doped sample, with x=15% and δ=21%, $d_{33}$ = 26±4 pm/V. This remarkably high value is comparable to the $d_{33}$ of PPLN - a device based on $LiNbO_3$, a widely used piezoelectric material (*27*) - and it is the highest reported or predicted $d_{33}$ value among the TMDs in any configuration (*36*, *37*). It is important to mention that previously reported piezoelectric properties in the TMD's have been measured in nanostructures of these compounds, generally in single layers in which centrosymmetry is broken, and recently, in thin films of TMD alloys (*38*). Our measurements are performed in macroscopic bulk single crystals, and as such, constitute, to the best of our knowledge, the first report of piezoelectricity in the bulk of TMDs, with piezoelectric coefficients that are comparable to the materials used in commercial piezoelectric devices.

In order to further characterize the electro-mechanical properties in these compounds, switching spectroscopy (SS)-PFM measurements were performed (see methods and section S5 for experimental details). The phase (figs. 3(e,g,i)) and amplitude (figs. 3(f,h,j)) of the SS-PFM piezoresponse were recorded as a function of the DC bias voltage at different random locations over the (insets to figures 3(e,g,i)). The (x=1.4%, δ=9%) Te-doped crystal (figs. 3(e,f)) reveals a piezoelectric response without hysteresis. Interestingly, the amplitude tends to saturate for $|V_{DC}|$>1V, suggesting a possible saturation of the electric dipole moment response. The (x=7.6%, δ=38%) and (x=15%, δ=21%) Te-doped samples show clear hysteresis loops both in the amplitude and phase of the SS-PFM piezoresponse (figs. 3(g,h) and 3(i,j), respectively), with shapes equivalent to the ones shown in prototypical ferroelectric materials such as $BaTiO_3$ and PZT (*39*). These loops indicate the presence of domains of electric polarization that can be aligned, with a coercive voltage of 0.2 V and 0.5V for the x=7.6% and 15% Te-doped bulk samples, respectively, and therefore reveal that these materials are ferroelectric.

In addition, electrical current vs voltage (IV) measurements taken in the Te-doped samples reveal asymmetric and hysteretic curves (fig. S8), with similar characteristics to those measured in the prototypical ferroelectric $BaTiO_3$ and multiferroics such as $BiFeO_3$ (*40*). This makes Te-doped $WSe_2$ suitable for a variety of multifunctional applications in electronic devices (*41*).

As presented in a previous section, the crystalline structure of our samples is the centrosymmetric 2H polytype. Previous reports of piezoelectric behavior in monolayer TMD's, and ferroelectric behavior in $WTe_2$ have attributed the non-centrosymmetric structures in those systems as the underlying mechanism for the existence of electric dipole moments. Therefore, these results invite us to search for alternative mechanisms for ferroelectricity in the TMD's, as will be presented in the computational modeling and discussion sections.

### Computational modeling results

In order to gain further insight into the mechanisms that can lead to the multiferroic properties of $W(Se_{1-x}Te_x)_{2(1-\delta)}$, first principles calculations were performed using density functional theory (DFT) (see methods). For each concentration random configurations were selected to perform the calculations. Resulting total magnetizations for different levels of Te doping and chalcogen vacancies can be found in Table S4, for one particular realizations of each composition. Interestingly, the system with the highest magnetic moment corresponds to the one with the highest value of chalcogen vacancies, $\delta$, independent of the value of Te doping, x.

Figure 4 shows different views of two layers of a simulated crystal structure with x≈10.26% and δ≈39%, very close to the values of one of our experimentally characterized samples that shows the strongest ferromagnetism and intermediate ferroelectricity. Arrows in figure 4(c) represent the magnetic moment contribution per atom for the atoms that contribute the most magnetic moment in this configuration. In this figure, in order to facilitate visualization, bonds between W atoms are drawn if the interatomic distance is equal or less than the bond distance in crystalline tungsten (2.74 Å); and bonds between W and chalcogen atoms are drawn if the interatomic distance is less than 2.72 Å and 2.747 Å for W-Se and W-Te bonds, respectively. The W atoms that are not bonded to a chalcogen element in the figure (labeled as W24 in fig. 4(a), and W1 and W5 in fig. 4(b)) are the ones that contribute the most magnetization of all atoms in the structure (as represented by the length of the gray arrows in fig. 4(c)). The next-higher contributions come from W atoms that have a reduced number of bonds with chalcogen atoms (this is, W atoms labeled as 17, 21, 23, 25, 28 and 29 in fig. 4(a), and 4, 6 and 16 in fig. 4(b), displayed as red arrows in fig. 4(c)). The magnitude of the contributions seems to be unaffected by whether the W is bonded to Te or Se atoms. Therefore, the relevant parameter that drives the magnitude of the magnetic moment is the number of bonds of each W with neighboring chalcogen atoms. Although the criteria for drawing a bond in figure 4 is rather arbitrary, a different number of bonds for each W ion is an indication of a different local environment for each. Therefore, our calculations indicate that the magnitude of the W magnetic moment in this system is driven by the local environment of the W atoms, and the strength of its hybridization to chalcogen atoms (or alternatively, its bonding to a chalcogen vacancy). Figure 4(d) shows the local spin density, $n_s(\mathbf{r})=n_u(\mathbf{r})-n_d(\mathbf{r})$, where $n_u$ ($n_d$) is the density of spin-up (spin-down) electrons, plotted on two planes containing W1, W5 and W24 (atoms without bonds to chalcogen atoms). It is important to notice that throughout the unit cell the spin density keeps low values ~±$5\times10^{-4}\mu_B$ (pink) while on these 3 atoms the spin density peaks at ~$8.8\times10^{-2}\mu_B$ (violet). For the atoms not bonded to a chalcogen element in figure 4(a-c), atomic magnetization (the integral of the spin density up to the covalent radius) takes values as high as 0.56 $\mu_B$. Furthermore, the overall ferromagnetic character of the system is corroborated by means of the total density of states (DOS) for spin-up and -down (fig. 4(e)), which shows that at the Fermi energy there is a larger contribution from the spin-up channel (red) than from the spin-down channel (blue). The projection of the DOS on atomic orbitals allows for the identification of tungsten d-orbitals as the electronic states associated with the found magnetic order, as shown in figs. 4(f) and S9(a).

In order to corroborate if magnetic order is dependent on the positions of chalcogen vacancies and/or Te substitutions, calculations were performed for a different configuration keeping the same concentration of vacancies and Te doping. In this second configuration (fig. S9(b)) there is a more homogeneous distribution of vacancies which causes that all W-atoms are bonded to at least one chalcogen-atom, either Se or Te. The resulting supercell magnetization is null in this case, with almost negligible atomic magnetizations (around $10^{-4}$ $\mu_B$) with opposite directions uniformly distributed throughout the cell.

In order to find possible sources for the ferroelectric state, electric polarization and ferroelectric behavior were analyzed in the framework of the modern theory of polarization (*42*), which allows to obtain the unit-cell polarization from the Berry phase of the system. The electric polarization vector was calculated in the same systems for which magnetization was studied. Results show a dependency on vacancy concentration and Te doping as can be seen in Table S5. Additionally, Hirshfeld charge analysis (*43*) was carried out in order to identify both the atoms responsible for polarization and the possible mechanism of ferroelectricity.

In the case of pristine $WSe_2$ it is found that chalcogen atoms have positive Hirshfeld charges while W atoms gain a negative charge which is approximately equal to twice the charge of the chalcogens. This is expected from the coordination numbers of Se and W, three and six respectively (Table S5). This distribution of charge makes each layer of 2H-$WSe_2$ non-polar due to its centrosymmetry, which is broken by the presence of vacancies. As can be seen in Figure 5, the absence of a Se atom develops electric polarization directed from the negative W plane towards the remaining Se atom below or above the vacancy. Since the sign of the polarization depends on the position of the chalcogen atom, this result suggests that polarization could be reverted by moving the chalcogen atom (or equivalently, the chalcogen vacancy) from one face of a layer to the other face, passing through the triangular space defined by the neighboring W atoms. This establishes a possible mechanism for the presence of switchable electric dipole moments in $WSe_2$. Interestingly, even though for this configuration there is a high concentration of vacancies ($\delta \approx 39\%$), there is only one place with a non-mirrored vacancy located at the top layer. The remaining vacancy sites are located on both sides of both layers of the unit cell, therefore not breaking its centrosymmetry, as can be seen in the bottom layers of the unit cell in figure 5.

To study the effect of Te doping on the electric polarization, Se5 (as labeled in Fig. 5) was exchanged by a Te atom. The resulting Hirshfeld charge of this Te is ~2.61 times larger than that of Se5 and the total polarization of the supercell presents a ~20 times enhancement if compared to the previous system. The larger separation of charge seen with Te can be ascribed to the difference in its electronegativity ($\chi$) with respect to both Tungsten and Selenium. In Pauling units $\chi^{Se}=2.55$, $\chi^W=2.36$ and $\chi^{Te}=2.1$, which accounts for the larger migration of charge from Te to W atoms. An additional configuration with a higher concentration of Te and less vacancies was studied ($x=16\%$ and $\delta=21.88\%$). In this case there are 13 non-mirrored vacancies, and chalcogen atoms are distributed on both faces of the two layers of the unit cell, 9 on the superior face and 4 in the inferior face. In general, Hirshfeld charges of Te range from 2 to 3 times the charges of Se, which gives larger local electric dipole moments. Since several local dipoles have opposite directions depending on the position of the vacancies, the computed unit cell polarization is about the same value found for the configuration with only one non-mirrored vacancy. In the case where all local dipoles point in the same direction, a large unit cell polarization is expected.

## 3. Discussion

Ferromagnetic states have been predicted theoretically (*31*) and experimentally realized in TMD through defects formation (*21*), vacancies formation (*23*), proton irradiation (*33*), nanostructuration (*34*), and doping with magnetic impurities ([22, 24]). In the case of this study, the dependency of the coercive fields on Te doping and chalcogen vacancies suggests that the magnetic properties of bulk $W(Se_{1-x}Te_x)_{2(1-\delta)}$, are driven by the amount of chalcogen vacancies, rather than by the presence of Te. This result aligns with the mechanism for ferromagnetic order derived from our DFT calculations, which reveal that the spin density is strongly localized onto the d-orbitals of the W-atoms less bonded with chalcogen atoms (or equivalently, more bonded to chalcogen vacancies), as presented in figures 4(f) and S9. This suggests that unpaired/localized d-electrons in W are responsible for the observed magnetic behavior. The key role of chalcogen vacancies in the magnetic order of TMDs has been previously predicted and experimentally observed in systems such as $MoS_2$ ([44]), $WS_2$ ([23]) and monolayer V-doped $WSe_2$ ([32]). For instance, the presence of vacancies in $2H\text{-}MoS_2$ nanosheets has been shown to transform the crystalline structure around the vacancies into a 1T-structure ([44]). In this environment the $Mo^{4+}$ ions have a net magnetic moment, in contrast to stoichiometric $2H\text{-}MoS_2$ in which the $Mo^{4+}$ ions are non-magnetic. Data suggests that the origin of the measured ferromagnetism in this system is intimately connected to an enhanced $Mo^{4+}$-S-vacancy exchange interaction. Similar conclusions have been recently reached in monolayer V-doped $WSe_2$, in which ferromagnetic properties are significantly enhanced by promoting Se-vacancies in the structure ([32]). In the case of $W(Se_{1-x}Te_x)_{2(1-\delta)}$, the role of chalcogen vacancies can be completely analogous to their TMD's counterparts. This is, a local magnetic moment is generated in the d-orbitals of transition metal atoms surrounded by vacancies due to their modified local environment. This can result in a strengthening of the exchange interaction through their coupling with chalcogen vacancies, leading to the ferromagnetic state.

On the other hand, ferroelectric states have been commonly associated with the non-centrosymmetry inherent to certain crystal structures. Among the TMD's, piezoelectric and ferroelectric properties have been previously reported in crystals with the non-centrosymmetric $1T_d$-phase, like the case of pure $WTe_2$ ([20]), or in nanostructures of the 2H-polytype with an odd number of layers, for which centrosymmetry is broken ([25, 26]). However, alternative mechanisms, other than the underlying bulk crystalline symmetry, have been proposed for the appearance of electric polarizations in ferroelectric materials. For example, in type-II multiferroics such as $TbMnO_3$ ([45]), $Ca_3CoMnO$ ([46]) and the 2D material $NiI_2$ ([14]), local electric polarization is proposed to be induced by the strong spin-orbit coupling in a magnetic host, which can lead to inversion symmetry breaking, and to a ferroelectric state. In addition, and for the specific case of TMD's, inequivalent Te-W distances in adjacent $1T\text{-}WTe_2$ layers has been theoretically shown to induce a charge imbalance, leading to an out-of-plane electric polarization ([47]). In the case of this work, our DFT calculations identify a feasible mechanism for the existence of electric dipole moments in the 2H unit cell that can lead to a non-zero total electric dipole moment in the crystal. In our model, the electric dipoles are originated in the charge imbalance created by a chalcogen vacancy aligned with a chalcogen atom in the c-direction. Our calculations indicate that such imbalance is an order of magnitude stronger if the chalcogen is a Te-atom, which is consistent with our experimental results: no ferroelectric state is found in the pure $WSe_{2(1-\delta)}$, in which, although the charge imbalance mechanism due to vacancies is also present, it is possibly not strong enough to result in a ferroelectric state. Whereas ferroelectricity grows stronger with Te doping, being well established for the largest Te-doped crystals.

For $W(Se_{1-x}Te_x)_{2(1-\delta)}$, the role of chalcogen vacancies seems to be crucial for both ferromagnetic and ferroelectric states. Interestingly, each of these ferroic orders in this material can exist independent from the other: $WSe_{2(1-\delta)}$ is only ferromagnetic and $WTe_2$ is only ferroelectric. This, together with the fact that multiferroicity in $W(Se_{1-x}Te_x)_{2(1-\delta)}$ is observed at room temperature, suggests that, although part of the origin of both ferroic orders can be common, this material is not a type-II multiferroic. In this class of multiferroics the mechanisms for both ferroic orders are highly intertwined through spin-orbit coupling, and generally show cryogenic critical temperatures (*45*). Nevertheless, the strong influence of chalcogen vacancies on both ferroic orders in $W(Se_{1-x}Te_x)_{2(1-\delta)}$ makes possible, not only the simultaneous presence of these states, but also their magnetoelectric coupling - a crucial ingredient in several envisaged applications of multiferroic materials. These ingredients, combined in an intrinsically 2D vdW layered material, and at room temperature, opens the door to a wide use of nanostructured multiferroic devices.

## 4. Methods

### Synthesis of single crystals

Tellurium-doped tungsten diselenide single crystals, $W(Se_{1-x}Te_x)_{2(1-\delta)}$ (with x=0-15% and δ=1-38%), were grown by chemical vapor transport (CVT) with iodine as transporting agent. Powders with the desired stoichiometry were produced from elemental W, Se and Te through a sintering process. For each batch, stoichiometric amounts of tungsten (No. 357421, Sigma Aldrich), selenium (No. 36208, Alfa Aesar) and tellurium (No. 266418, Sigma Aldrich) were thoroughly ground, and then cold-pressed at 1000 psi to form a pellet. The pellet was sealed in a quartz tube, leaving a small pressure of argon inside the quartz ampoule. The ampoule was then heated from room temperature to 500°C in 24h and maintained at this temperature for 72h. The obtained sintered powder and the iodine were then encapsulated in vacuum in a long quartz tube, and then placed in a two-zone tube furnace with 1010°C in the hot-zone and 900°C in the cold-growth-zone, for 144 hours. The single crystals were collected from the growth-zone of the tubes.

### X-ray diffraction

The doping values x and δ of our $W(Se_{1-x}Te_x)_{2(1-\delta)}$ single crystals were determined by triplication in X-ray fluorescence (XRF) using a ZSX Primus RIGAKU spectrometer.

X-ray powder diffraction patterns of $W(Se_{1-x}Te_x)_{2(1-\delta)}$ single crystals for all compositions were obtained using an Empyrean PANalytical series 2 diffractometer with a CuKα radiation source (λ = 1.5405980 Å), operated at 40 mA, 45kV, and with a step size of 0.0262606° over a 2θ range from 10° to 70°, in an Eulerian-Cradle geometry. This characterization was performed in a collection of single crystals of each batch, which were cut in tiny fragments and spread in the sample holder to emulate a powder diffraction experiment. This was done with the purpose of obtaining a statistically meaningful characterization of many crystals of each batch. The most intense diffraction peaks for all batches come from the {0 0 1} family of planes, given that the *a-b* plane of the crystals tend to align parallel to the sample holder. Few peaks corresponding to different families of planes can still be recognized.

In order to resolve the exact crystal structures of the x=1.4% and 7.6% Te-doped samples, single crystal X-ray diffraction was also measured. The intensities were measured at room temperature, 298 (2) K, using CuKα radiation (λ = 1.54184 Å), and ω scans in an Agilent SuperNova, Dual, Cu at Zero, Atlas four-circle diffractometer equipped with a CCD plate detector. The collected frames were integrated with the CrysAlis PRO software package (CrysAlisPro 1.171.39.46e, Rigaku

Oxford Diffraction, 2018). Data were corrected for the absorption effect using the CrysAlis PRO software package by the empirical absorption correction using spherical harmonics, implemented in the SCALE3 ABSPACK scaling algorithm (CrysAlisPro 1.171.39.46e, Rigaku Oxford Diffraction, 2018). The structures were solved using an iterative algorithm (*48*), and then completed by difference Fourier map. The crystal structures were refined using the program SHELXL2018/3 (*49*).

### Raman spectroscopy

Raman spectra for crystals of all the compositions studied were taken in a HORIBA Scientific XPLORA Xl041210 Raman spectrometer. The excitation wavelength used for all measurements was 532 nm with a grating of 2400 lines/mm, in the range of 70 $cm^{-1}$ to 1000 $cm^{-1}$.

### Magnetic measurements

Isothermal magnetic hysteresis loops at 300K and 80K were measured in a Lakeshore[TM] 7400 Series vibrating-sample magnetometer (VSM). Cryogenic temperatures were reached through a flow-cryostat operated with liquid nitrogen.

### Piezoresponse Force Microscopy Measurements

The piezoelectric effect can be measured through piezoresponse force microscopy (PFM) technique, which is a variation of the atomic force microscopy (AFM) technique. The piezoresponse is acquired when an AC electrical voltage is applied using a conductive tip in contact with the sample surface. The resulting oscillatory deformation of the sample under the AC voltage is detected through the AFM cantilever deflection. We use two kinds of PFM measurement: Contact resonance (Dual AC Resonance Tracking- DART) and Switching Spectroscopy - SS-PFM -, using a Cypher ES Environmental AFM operated at room temperature and in an inert $N_2$ atmosphere.

In the DART-PFM mode the cantilever is operated with a AC voltage frequency near the contact resonance frequency, resulting in an driven harmonic oscillator, which enhances the piezoresponse signal, even if it is very small. Similarly, SS-PFM uses a sinusoidal voltage at contact resonance frequency overlapped with square voltages of smaller periods. In ferroelectric materials, it allows the determination of the electromechanical hysteresis loops and their switching parameters such as coercive and nucleation voltages.

To avoid misunderstandings in piezoresponse amplitude data, arbitrary units (a. u.) are used when piezoresponse amplitude is reported without treatment, whereas picometers (pm) are used when amplitude data is normalized by a single harmonic oscillator (SHO) model (*50*). This model provides an accurate way to calculate the real piezoelectric amplitude in the small damping regime. Moreover, to demonstrate the accuracy of this model and its use in a methodology to obtain the effective piezoelectric constant, $d_{33}$ in our single crystals, an standard sample of periodically poled lithium niobate (PPLN) was characterized by DART-PFM and SS-PFM using silicon probes (AC240TM-R3) (sections S4 and S5 for more details). Afterwards, the same methodology was used in our doped single crystals (section S5 for data without data analysis). Experimental spring constants were between 1.71-2.95N/m, and contact-mode resonance frequencies were between 219-241kHz.

### Transport properties

Electrical properties of single crystals were measured by a four-probe technique for resistance vs temperature/field measurements, and two-probe technique for current vs voltage measurements, using a Keithley 2400 Source Meter. Gold pads were evaporated into the single crystals in order to reduce contact resistance and capacitive effects. Cryogenic temperatures were achieved using an Oxford Instruments liquid helium variable temperature insert cryostat equipped in an IntegraAC Recondensing Helium Magnet System.

### DFT calculations

First principles calculations were performed using Density Functional Theory (*51*) (DFT) as implemented in the Quantum ESPRESSO package (*52*). Core electrons were treated within the pseudopotential approximation by means of norm-conserving ultrasoft pseudopotentials. Generalized gradient approximation (GGA) in the parametrization of Perdew-Burke-Ernzerhof (PBE) (*53*), was chosen for the exchange-correlation (XC) energy functional. Grimme's semiempirical DFT-D3 method (*54*) was used to include van der Waals interactions between the layers, ignoring three-body terms. The number of plane-waves in the basis set correspond to a kinetic energy cutoff of 50 Ha. Structural optimizations were performed until forces reached a tolerance of $1x10^{-4}$ a.u.

Starting with the simulation of pure $WSe_2$, ionic positions and unit cell lattice parameters were optimized. Resulting structural parameters are in good agreement with the measured ones and with previously reported DFT calculations (*62*), as can be seen in table S4 of section S7. Inclusion of both, chalcogen vacancies and Te-doping, was achieved in a 4x4x1 supercell, which allowed the simulation of systems with Te-doping and chalcogen vacancy levels similar to the measured single crystals. The effect of vacancies and Te doping on the magnetic behavior was analyzed by performing spin-polarized calculations with and without spin-orbit coupling. For each concentration a random configuration was selected to perform the calculations. In all cases a 4 x 4 x 3 Monkhorst-Pack grid (*55*) was used for sampling the first Brillouin zone.

Berry phase calculations of electric polarization were performed with the implementation found in the ABINIT(*56*) program using a finer reciprocal space grid of 5 x 5 x 3 points and ONCVPSP (*57*) pseudopotentials. Hirshfeld charge analysis was performed with the post-processing tool *cut3d* from ABINIT.

## Acknowledgements

**Funding:** The authors thank A. Castellanos-Gómez and C. Munuera for enlightening discussions, and Daniel Dorado for his contribution to IV curves measurements. G.C-C., K.V-B., J.P., O.H., C.E., W.L.P., J.A.G. and P.G-G. thank the financial support of the Ministerio de Ciencia, Tecnología e Innovación de Colombia, through Grant No. 122585271058. C.E. and W.L.P. acknowledge computational resources from Granado-HPC, Universidad del Norte. P.G-G. and K.V-B. thank the support of the School of Sciences and the Vice Presidency of Research Creation at Universidad de Los Andes. P.G-G., G.C-C. and K.V-B. thank the financial support of FPIT - Fundación para la promoción de la investigación y la tecnología, of Banco de la República de Colombia, Project number 4.687. J.A.G., J.P. and O.H. thank the support of Clúster de Investigación en Ciencias y Tecnologías Convergentes NBIC, Universidad Central. P. G-G. and D. S-V. thank the support of the Consejo Superior de Investigaciones Científicas (CSIC) of Spain, through the i-COOP+ 2020 project, #COOPA20460.

and secured funds. All authors participated in the discussion of the data and manuscript writing.

**Competing Interests:** The authors declare no competing financial interests.

**Data and materials availability:** All data related to this work is available upon reasonable request to the corresponding author.

*References*

1.  J.-M. Hu, et al. , Multiferroic magnetoelectric nanostructures for novel device applications. *MRS Bull.* **40**, 728–735 (2015).

2.  W. Eerenstein, et al. , Multiferroic and magnetoelectric materials. *Nature*. **442**, 759–765 (2006).

3.  H. Béa, et al. , Spintronics with multiferroics. *J. Phys. Condens. Matter*. **20**, 434221 (2008).

4.  J. F. Scott, Multiferroic memories. *Nat. Mater.* **6** (2007), pp. 256–257.

5.  N. A. Hill, Why Are There so Few Magnetic Ferroelectrics? *J. Phys. Chem. B*. **104**, 6694–6709 (2000).

6.  C. Chen, et al. , Strong d–d electron interaction inducing ferromagnetism in Mn-doped LiNbO3. *Thin Solid Films*. **520**, 764–768 (2011).

7.  F. Pan, et al. , Ferromagnetism and possible application in spintronics of transition-metal-doped ZnO films. *Mater. Sci. Eng. R Rep.* **62**, 1–35 (2008).

8.  E. Hanamura, Y. Tanabe, Phase transitions and second-harmonics of ferroelectric and antiferromagnetic RMnO3. *Phase Transitions*. **79**, 957–971 (2006).

9.  J. Wang, et al. , Epitaxial BiFeO3 multiferroic thin film heterostructures. *Science*. **299**, 1719–1722 (2003).

10. R. Ramesh, N. A. Spaldin, Multiferroics: progress and prospects in thin films. *Nat. Mater.* **6**, 21–29 (2007).

11. X. Tang, L. Kou, Two-Dimensional Ferroics and Multiferroics: Platforms for New Physics and Applications. *J. Phys. Chem. Lett.* **10**, 6634–6649 (2019).

12. B. Behera, B. C. Sutar, N. R. Pradhan, Recent progress on 2D ferroelectric and multiferroic materials, challenges, and opportunity. *Emergent Materials*. **4**, 847–863 (2021).

13. Y. Gao, M. Gao, Y. Lu, Two-dimensional multiferroics. *Nanoscale*. **13**, 19324–19340 (2021).

14. Q. Song, et al. , Evidence for a single-layer van der Waals multiferroic. *Nature*. **602**, 601–605 (2022).

15. Y. Zhao, et al. , Surface Vacancy-Induced Switchable Electric Polarization and Enhanced Ferromagnetism in Monolayer Metal Trihalides. *Nano Lett.* **18**, 2943–2949 (2018).

16. H. Yang, el al. , Iron-doping induced multiferroic in two-dimensional In2Se3. *Science China Materials*. **63**, 421–428 (2019).

17. T. Zhong, et al. , Room-temperature multiferroicity and diversified magnetoelectric couplings in 2D materials. *Natl Sci Rev*. **7**, 373–380 (2020).

18. Z. Shao, et al. , Multiferroic materials based on transition-metal dichalcogenides: Potential platform for reversible control of Dzyaloshinskii-Moriya interaction and skyrmion via electric field. *Phys. Rev. B Condens. Matter*. **105**, 174404 (2022).

19. M. Bonilla, et al. , Strong room-temperature ferromagnetism in VSe2 monolayers on van der Waals substrates. *Nat. Nanotechnol.* **13**, 289–293 (2018).

20. P. Sharma, et al. , A room-temperature ferroelectric semimetal. *Sci Adv*. **5**, eaax5080 (2019).

21. S. Tongay, et al. , Magnetic properties of MoS2: Existence of ferromagnetism. *Appl. Phys. Lett.* **101**, 123105 (2012).

22. M. Habib, et al. , Ferromagnetism in CVT grown tungsten diselenide single crystals with

nickel doping. *Nanotechnology*. **29**, 115701 (2018).

23. X. Ding, et al. , Enhanced ferromagnetism in WS2 via defect engineering. *J. Alloys Compd.* **772**, 740–744 (2019).

24. Y. T. H. Pham, et al. , Tunable Ferromagnetism and Thermally Induced Spin Flip in Vanadium-Doped Tungsten Diselenide Monolayers at Room Temperature. *Adv. Mater.* **32**, e2003607 (2020).

25. E. Nasr Esfahani, et al. , Piezoelectricity of atomically thin WSe2 via laterally excited scanning probe microscopy. *Nano Energy*. **52**, 117–122 (2018).

26. W. Wu, et al. , Piezoelectricity of single-atomic-layer MoS2 for energy conversion and piezotronics. *Nature*. **514**, 470–474 (2014).

27. R. S. Weis, T. K. Gaylord, Lithium niobate: Summary of physical properties and crystal structure. *Appl. Phys. A: Mater. Sci. Process.* **37**, 191–203 (1985).

28. J.-H. Lee, et al. , Reliable Piezoelectricity in Bilayer WSe2 for Piezoelectric Nanogenerators. *Adv. Mater.* **29** (2017), doi:10.1002/adma.201606667.

29. 2H-WSe2 (WSe2) crystal structure, (available at https://materials.springer.com/isp/crystallographic/docs/sd_0310430).

30. P. Yu, et al. , Metal-Semiconductor Phase-Transition in WSe2(1-x)Te2x Monolayer. *Adv. Mater.* **29** (2017), doi:10.1002/adma.201603991.

31. Z. Zhang, et al. , Intrinsic magnetism of grain boundaries in two-dimensional metal dichalcogenides. *ACS Nano*. **7**, 10475–10481 (2013).

32. S. J. Yun, et al. , Escalating Ferromagnetic Order via Se-Vacancies Near Vanadium in WSe Monolayers. *Adv. Mater.* **34**, e2106551 (2022).

33. S. W. Han, et al. , Controlling ferromagnetic easy axis in a layered MoS2 single crystal. *Phys. Rev. Lett.* **110**, 247201 (2013).

34. L. Tao, et al. , Experimental and theoretical evidence for the ferromagnetic edge in WSe2 nanosheets. *Nanoscale*. **9**, 4898–4906 (2017).

35. K.-A. N. Duerloo, et al. , Intrinsic Piezoelectricity in Two-Dimensional Materials. *J. Phys. Chem. Lett.* **3**, 2871–2876 (2012).

36. L. Rogée, et al. , Ferroelectricity in untwisted heterobilayers of transition metal dichalcogenides. *Science*. **376**, 973–978 (2022).

37. A. Rawat, et al. , Nanoscale Interfaces of Janus Monolayers of Transition Metal Dichalcogenides for 2D Photovoltaic and Piezoelectric Applications. *J. Phys. Chem. C.* **124**, 10385–10397 (2020).

38. Y. Chen, et al. , 2D Transition Metal Dichalcogenide with Increased Entropy for Piezoelectric Electronics. *Adv. Mater.*, e2201630 (2022).

39. S. Hong, et al. , Principle of ferroelectric domain imaging using atomic force microscope. *J. Appl. Phys.* **89**, 1377–1386 (2001).

40. F. Yan, G. Z. Xing, L. Li, Low temperature dependent ferroelectric resistive switching in epitaxial BiFeO3 films. *Appl. Phys. Lett.* **104**, 132904 (2014).

41. A. Sawa, Resistive switching in transition metal oxides. *Mater. Today*. **11**, 28–36 (2008).

42. R. Resta, Macroscopic Electric Polarization as a Geometric Quantum Phase. *EPL*. **22**, 133 (1993).

43. F. L. Hirshfeld, Bonded-atom fragments for describing molecular charge densities. *Theor. Chim. Acta*. **44**, 129–138 (1977).

44. R. Shidpour, M. Manteghian, A density functional study of strong local magnetism creation on MoS2 nanoribbon by sulfur vacancy. *Nanoscale*. **2**, 1429–1435 (2010).

45. M. Matsubara, et al. , Multiferroics. Magnetoelectric domain control in multiferroic TbMnO₃. *Science*. **348**, 1112–1115 (2015).

46. H. Wu, et al. , Ising magnetism and ferroelectricity in Ca3CoMnO6. *Phys. Rev. Lett.* **102**, 026404 (2009).

47. Z. Fei, et al. , Ferroelectric switching of a two-dimensional metal. *Nature*. **560**, 336–339 (2018).

48. L. Palatinus, G. Chapuis, SUPERFLIP – a computer program for the solution of crystal structures by charge flipping in arbitrary dimensions. *J. Appl. Crystallogr.* **40**, 786–790

(2007).

49. G. M. Sheldrick, Crystal structure refinement with SHELXL. *Acta Crystallogr. B.* **71**, 3–8 (2015).

50. R. García, R. Pérez, Dynamic atomic force microscopy methods. *Surf. Sci. Rep.* **47**, 197–301 (2002).

51. W. Kohn, L. J. Sham, Self-Consistent Equations Including Exchange and Correlation Effects. *Phys. Rev.* **140**, A1133–A1138 (1965).

52. P. Giannozzi, et al. , QUANTUM ESPRESSO: a modular and open-source software project for quantum simulations of materials. *J. Phys. Condens. Matter.* **21**, 395502 (2009).

53. J. P. Perdew, K. Burke, M. Ernzerhof, Generalized Gradient Approximation Made Simple. *Phys. Rev. Lett.* **77**, 3865–3868 (1996).

54. S. Grimme, J. Antony, S. Ehrlich, H. Krieg, A consistent and accurate ab initio parametrization of density functional dispersion correction (DFT-D) for the 94 elements H-Pu. *J. Chem. Phys.* **132**, 154104 (2010).

55. H. J. Monkhorst, J. D. Pack, Special points for Brillouin-zone integrations. *Phys. Rev. B Condens. Matter.* **13**, 5188–5192 (1976).

56. X. Gonze, et al. , The Abinit project: Impact, environment and recent developments. *Comput. Phys. Commun.* **248**, 107042 (2020).

57. D. R. Hamann, Optimized norm-conserving Vanderbilt pseudopotentials. *Phys. Rev. B Condens. Matter.* **88**, 085117 (2013).

58. L.-P. Feng, et al. , Effect of pressure on elastic, mechanical and electronic properties of WSe2: A first-principles study. *Mater. Res. Bull.* **50**, 503–508 (2014).

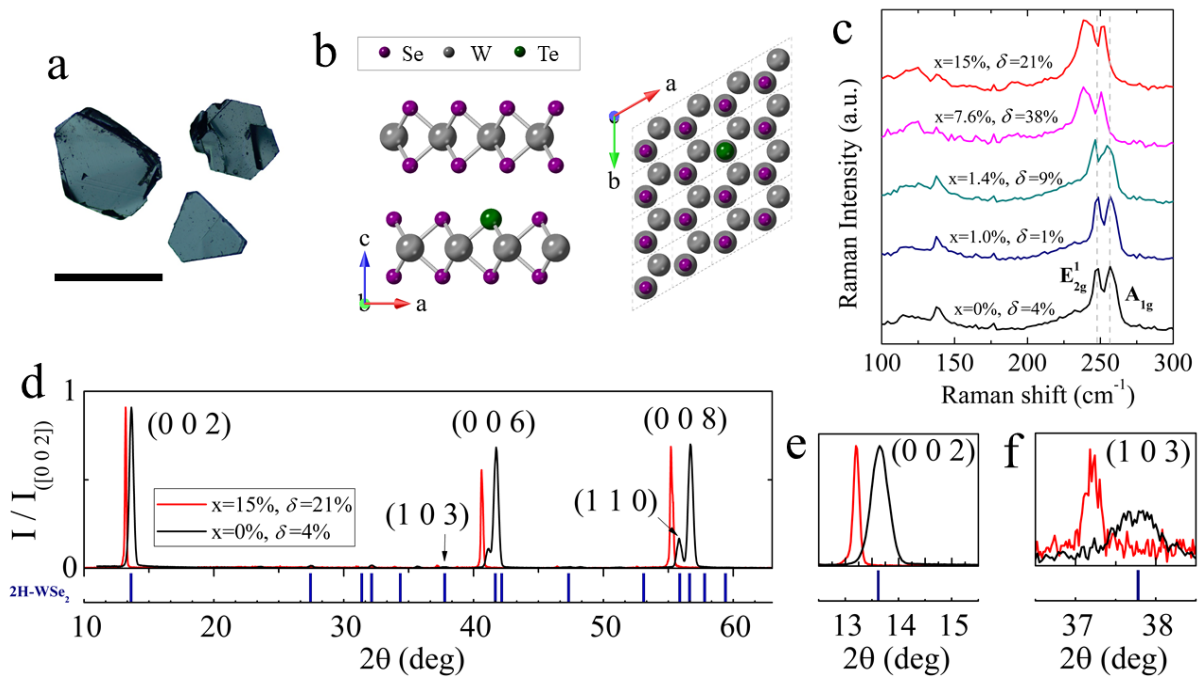

**Figure 1. Crystalline Structure. (a)** Representative single crystals of W(Se$_{1-x}$Te$_x$)$_{2(1-\delta)}$ (this case, for x=15% and $\delta$ =21%) synthesized by chemical vapor transport, and characterized in this work. Scale bar is 1 mm long. **(b)** Crystal structure of a 2H-polytype of Te doped WSe$_2$ with space group P6$_3$/mmc (#194). **(c)** Raman spectroscopy measurements for W(Se$_{1-x}$Te$_x$)$_{2(1-\delta)}$ single crystals of the compositions studied in this work. Data shows a close up look to the Raman shift around the characteristic E$_{2g}$ and A$_{1g}$ peaks of the 2H phase, which shift to lower values with increasing doping. Dashed gray lines indicate the peak positions for the pure x=0 compound. **(d)** Powder X-ray diffraction data for a collection of undoped (x=0, $\delta$=4%), and highest Te-doped (x=15%, $\delta$=21%) samples. Diffraction peaks are consistent with the 2H polytype, as compared with the reported peak positions of pure 2H-WSe$_2$, shown by the blue vertical lines (*29*). **(e, f)** Close-up look to the (0 0 2) and (1 0 3) peaks, which reveal a shift to the left for the Te-doped samples, indicative of an increase of 2.5% in the *c* lattice parameter

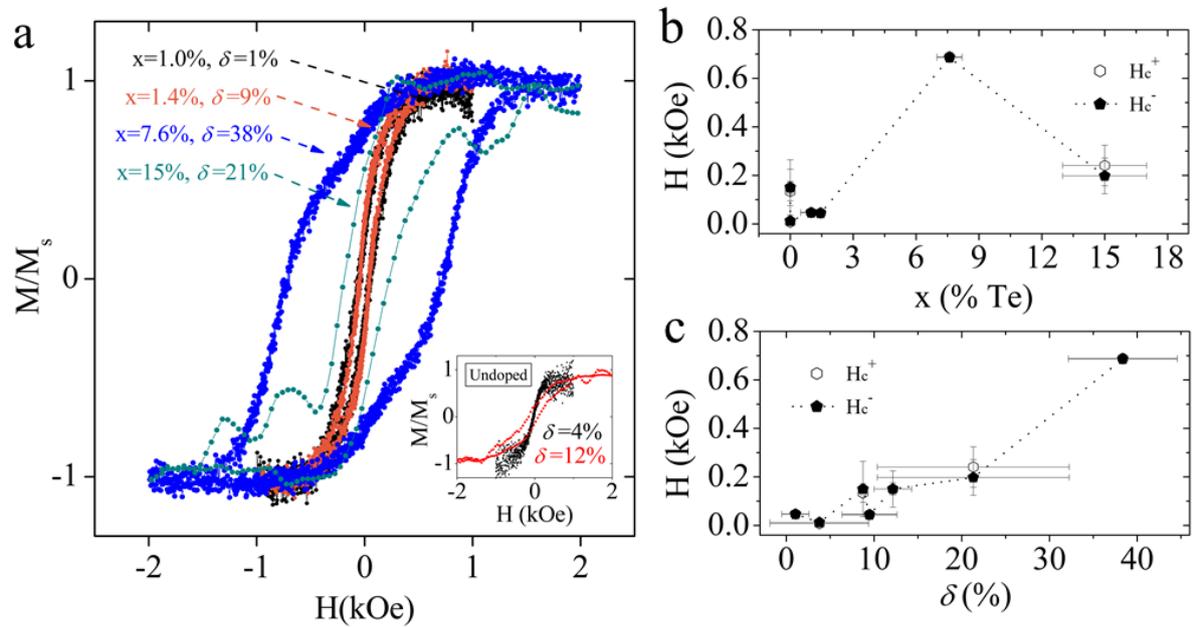

**Figure 2. Magnetic properties. (a)** Magnetization hysteresis loops at 300 K for samples of W(Se$_{1-x}$Te$_x$)$_{2(1-\delta)}$ with different x and $\delta$ values. For all curves, the diamagnetic components were subtracted, and their values normalized by the saturation magnetization, M$_s$. Inset in (a) shows the hysteresis loops for undoped (x=0) WSe$_{2(1-\delta)}$ samples with different amounts of chalcogen vacancies, $\delta$. **(b)** and **(c)** show coercive fields H$_c$ for all samples presented in (a), as a function of (b) Te-doping, x, and (c) chalcogen vacancies, $\delta$.

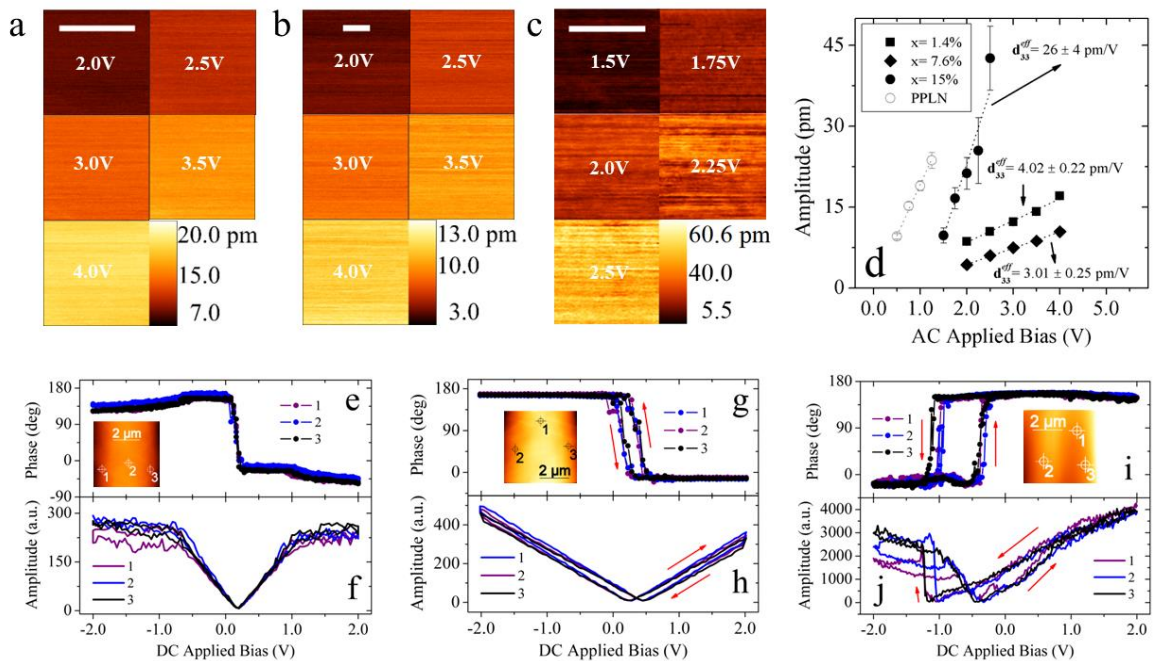

**Figure 3. Ferroelectric properties.** DART-PFM piezoresponse amplitude normalized by SHO model as function of the AC applied bias for $W(Se_{1-x}Te_x)_{2(1-\delta)}$ samples with **(a)** x=1.4%, δ=9%; **(b)** x=7.6%, δ=38%; and **(c)** x=15%, δ=21%. Scale bars of all three figures are 200 nm long. **(d)** Spatial average of the DART-PFM piezoresponse values from (a), (b) and (c) versus AC applied bias. The effective $d_{33}$ piezoelectric coefficient is obtained from the slope of these curves. The curve for a sample of PPLN is included for comparison purposes. SS-PFM piezoresponse phase and amplitude for samples with **(e-f)** x=1.4%, δ=9%; **(g-h)** x=7.6%, δ=38%; and **(i-j)** x=15%, δ=21%. Insets to all phase curves show topography images indicating the exact locations at which the SS-PFM piezoresponse curves were performed.

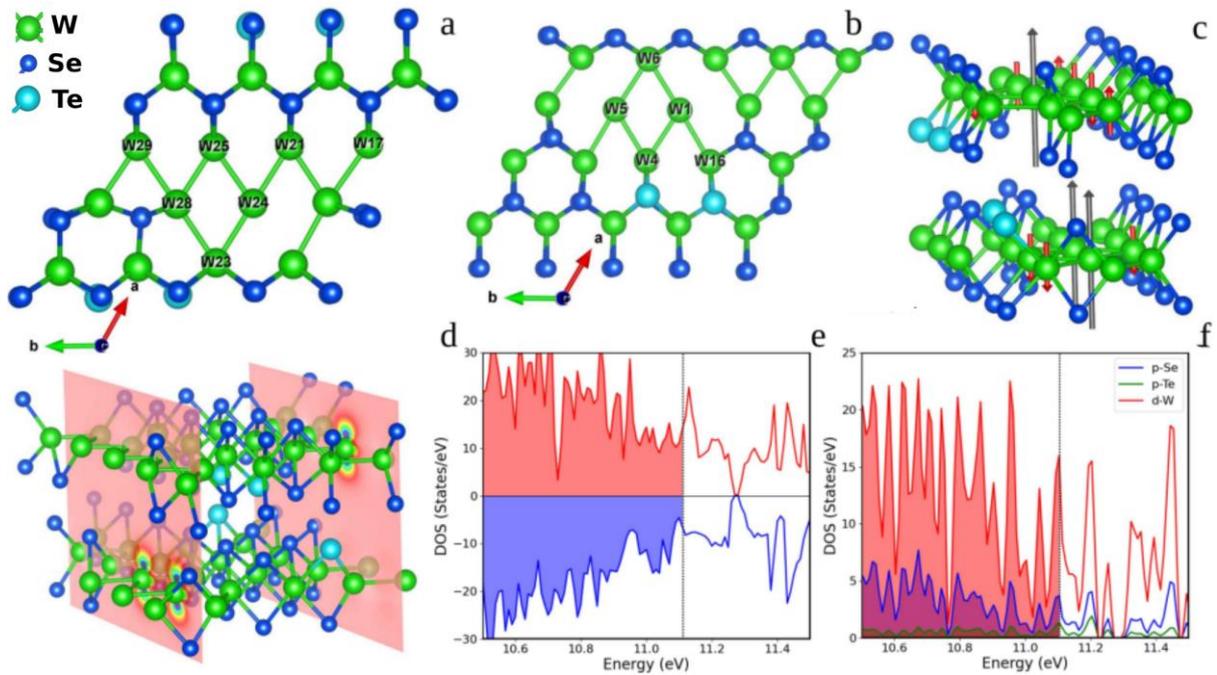

**Figure 4. Structural and magnetic calculations** of $W(Se_{1-x}Te_x)_{2(1-\delta)}$ for x=10.26% and δ=39.06%. W-W and W-Te bonds are drawn if the interatomic distances are less than 2.74 Å and 2.81 Å, respectively (see text). Top views of the **(a)** superior and **(b)** inferior unit cell monolayers and **(c)** perspective view of the optimized structure. Arrows indicate resulting atomic magnetic moments more than 0.02 $\mu_B$ in magnitude. **(d)** Spin density plotted on planes containing atoms W1, W5 and W24. To facilitate visualization, atom W24 from a neighboring unit cell is depicted. **(e)** Total spin-up (red) and spin-down (blue) density of states. **(f)** Projected density of states on p orbitals of Se and Te atoms, and on d orbitals of W.

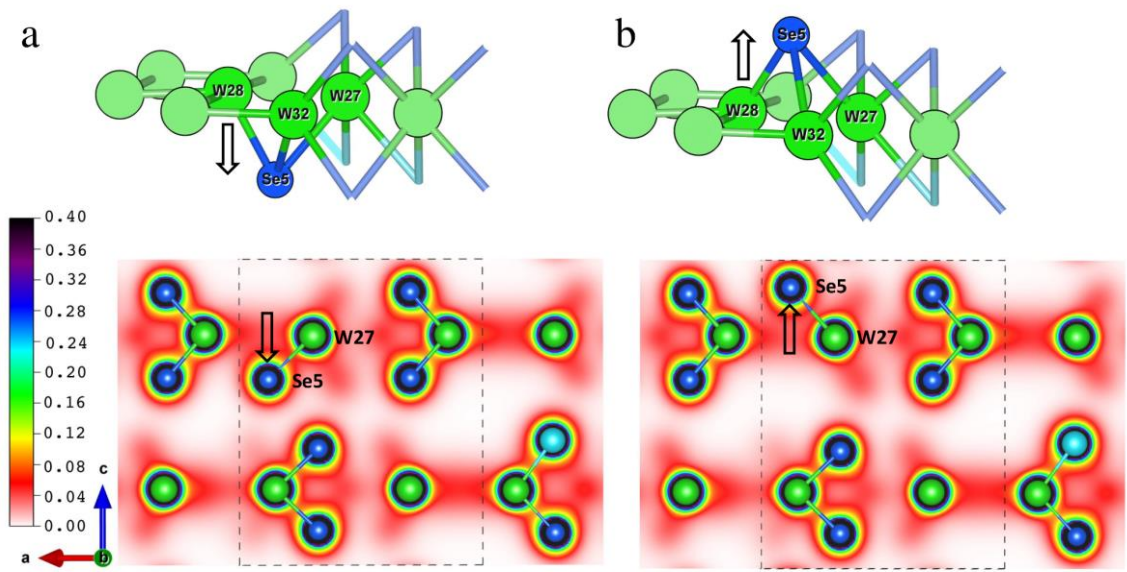

**Figure 5. Electric polarization calculations. (a)** Different views of a (0 1 0) plane of a $W(Se_{1-x}Te_x)_{2(1-\delta)}$ simulated structure with x=10.26% and δ=39.06%. The top view is a 3D-perspective view of the atoms around a chalcogen vacancy located in the upper side of the top-layer of the unit cell. The bottom view shows both layers of the unit cell (marked in the dashed square), and the color scale represents a slice of the total electronic density (color bar scale in a.u.). Arrows indicate the direction of the electric dipole moment around the vacancy, in the c-direction. **(b)** Same representations as in (a), but now with the chalcogen vacancy located in the bottom side of the top layer of the unit cell. The direction of the electric dipole moment is flipped in this case, as indicated by the arrow.

# Supplementary Materials

## Room temperature multiferroicity in a transition metal dichalcogenide

G. Cardenas-Chirivi[1,2], K. Vega-Bustos[1], H. Rojas-Páez[1], D. Silvera-Vega[1], J. Pazos[2], O. Herrera[2], M. A. Macías[3], C. Espejo[4], W. López-Pérez[4], J. A. Galvis[5] and P. Giraldo-Gallo[1]

[1] Department of Physics, Universidad de Los Andes, Bogotá 111711, Colombia
[2] Faculty of Engineering and Basic Sciences, Universidad Central, Bogotá, Colombia
[3] Department of Chemistry, Universidad de Los Andes, Bogotá 111711, Colombia
[4] Department of Physics and Geosciences, Universidad del Norte, Barranquilla, Colombia
[5] School of Engineering, Science and Technology, Universidad del Rosario, Bogotá 111711, Colombia

## Section S1. Chemical composition of single crystals

Chemical compositions for all single crystals were determined through XRF measurements, by triplication. According to the stoichiometric formula, $W(Se_{1-x}Te_x)_{2(1-\delta)}$, percentages of tellurium doping and chalcogen vacancies are presented in Table S1. The "Nominal Composition" column presents the aimed Te-doping and chalcogen vacancies concentration, used to calculale the stoichiometric fractions of elemental precursors during the initial sintering process. However, the real compositions, determined through XRF, differ considerably from the nominal values, as shown in the last two columns (real concentrations of Te and vacancies).

Several Te-doping concentrations were obtained. In addition, two undoped compositions with a high-level of vacancies (x=0%, δ=50%) were synthesized in order to discriminate the effects of Te-doping and chalcogen vacancies in the observed properties.

| Nominal composition | | Element | | | Real tellurium doping, x (%) | Real chalcogen vacancies, δ (%) |
|---|---|---|---|---|---|---|
| | | W | Se | Te | | |
| **x=0% δ=0%** | Mass (%) | 55 ± 2 | 45 ± 2 | - | 0 | 4 ± 6 |
| | moles | 1.00 ± 0.03 | 1.9 ± 0.1 | - | | |
| **x=0% δ=50%** | Mass (%) | 56.1 ± 0.1 | 43.9 ± 0.1 | - | 0 | 8.7 ± 0.2 |
| | moles | 1.000 ± 0.001 | 1.824 ± 0.005 | - | | |
| **x=0% δ=50%** | Mass (%) | 56.99 ± 0.84 | 43.01 ± 0.84 | - | 0 | 12 ± 3 |
| | moles | 1.00 ± 0.01 | 1.75 ± 0.04 | - | | |
| **x=2%** **δ=0%** | Mass (%) | 53.82 ± 0.08 | 45.3 ± 0.5 | 0.7 ± 0.4 | 1.0 ± 0.5 | 1 ± 2 |
| | moles | 1.000 ± 0.001 | 1.96 ± 0.02 | 0.02 ± 0.01 | | |
| **x=5%** **δ=0%** | Mass (%) | 56 ± 1 | 43 ± 1 | 1.08 ± 0.06 | 1.45 ± 0.08 | 9 ± 3 |
| | moles | 1.00 ± 0.02 | 1.78 ± 0.06 | 0.028 ± 0.002 | | |
| **x=50%,** **δ=0%** | Mass (%) | 64 ± 3 | 31 ± 3 | 5.4 ± 0.3 | 7.6 ± 0.6 | 38 ± 6 |
| | moles | 1.00 ± 0.04 | 1.1 ± 0.1 | 0.12 ± 0.01 | | |
| **x=50%,** **δ=0%** | Mass (%) | 57 ± 3 | 32 ± 4 | 11 ± 1 | 15 ± 2 | 21 ± 11 |
| | moles | 1.00 ± 0.04 | 1.3 ± 0.2 | 0.27 ± 0.04 | | |

**Table S1.** Chemical composition of single crystals, determined through X-ray Fluorescence Spectroscopy (XRF).

## Section S2. Raman and XRD spectra

In fig. S1, Raman spectra are shown for all Te-doped compositions and an undoped one, for a wide range of Raman shifts. A decrease in the Raman shift of the $E_{2g}$ and $A_{1g}$ modes of the 2H-structure, with increasing Te-concentration is observed.

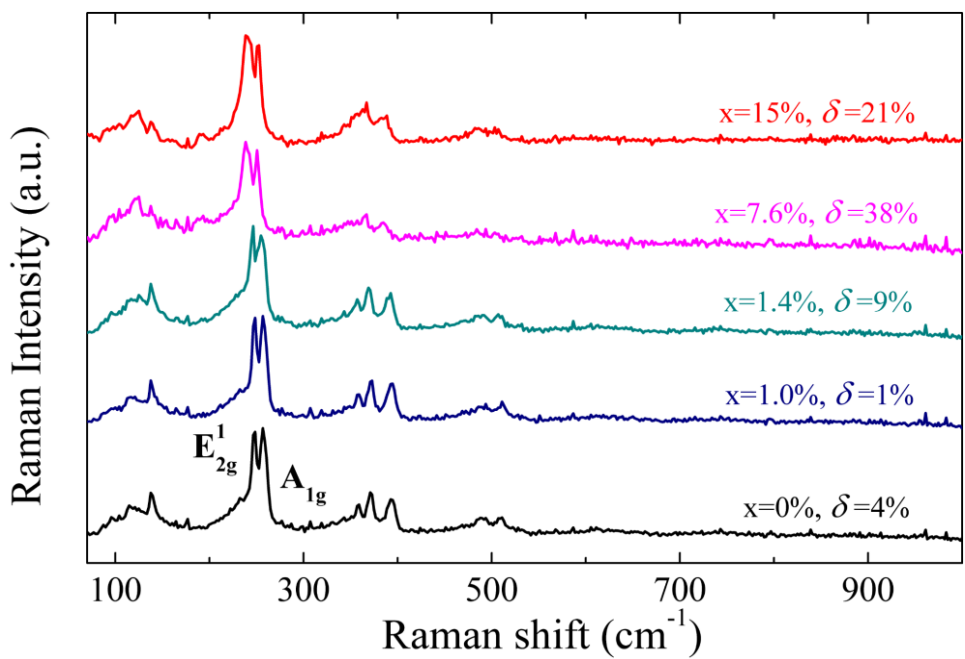

**Figure S1.** Raman spectra for all tellurium-doped single crystals involved in this study.

On the other hand, X-ray powder diffraction (XRD) patterns of $W(Se_{1-x}Te_x)_{2(1-\delta)}$ single crystals for all doped compositions and an undoped one are presented in fig. S2. Here, reported XRD peak positions for pure $2H\text{-}WSe_2$ (29) are added for comparison. A continuous shift to lower angle is observed in the {001} family peak positions with increasing Te-doping. This allows us to conclude that inclusion of tellurium in the 2H-phase enhances the c-lattice parameter with respect to the undoped one.

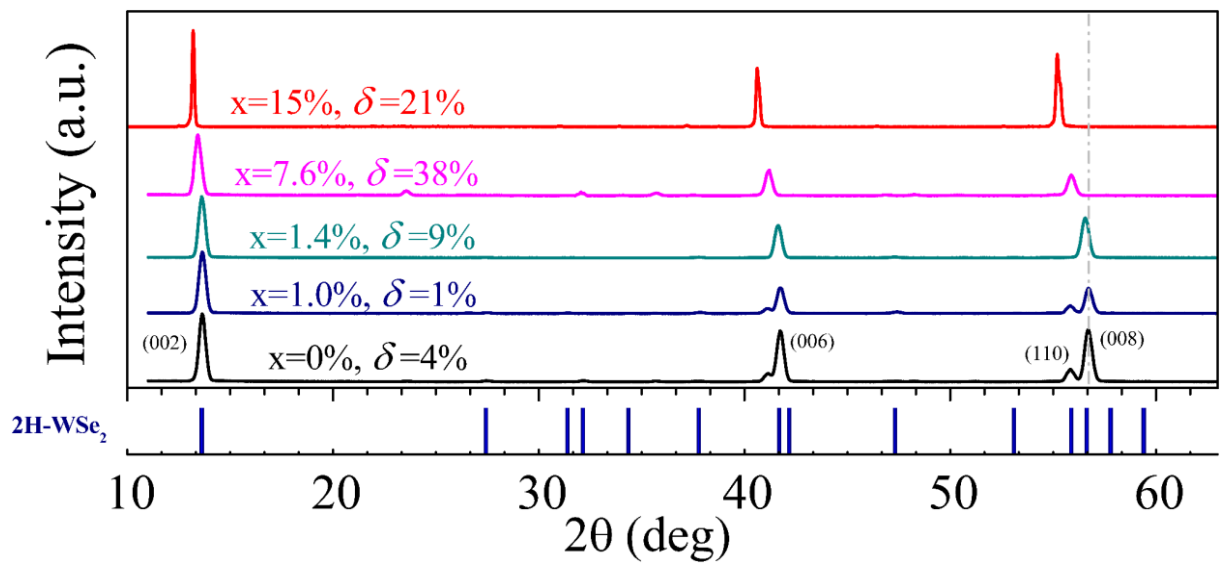

**Figure S2.** XRD patterns for all tellurium-doped single crystals involved in this study.

Table S2 presents the obtained lattice constants, obtained through powder x-ray diffraction and single crystal diffraction, as explained in the methods section. For the powder x-ray diffraction, the values shown in the table are the average over different peaks of the same family.

| Tellurium doping (x, %) | Powder X-ray Diffraction | | Single crystal X-ray Diffraction | |
|---|---|---|---|---|
| | $a$ (Å) | $c$ (Å) | $a$ (Å) | $c$ (Å) |
| 0 | $3.287 \pm 0.008$ | $12.974 \pm 0.009$ | - | - |
| 1.0 | $3.28 \pm 0.01$ | $12.99 \pm 0.03$ | $3.2857 \pm 0.0005$ | $12.9773 \pm 0.0015$ |
| 1.4 | $3.288 \pm 0.003$ | $13.004 \pm 0.008$ | - | - |
| 7.6 | $3.28 \pm 0.01$ | $13.16 \pm 0.02$ | $3.3074 \pm 0.0005$ | $13.180 \pm 0.002$ |
| 15 | $3.34 \pm 0.01$ | $13.3 \pm 0.1$ | - | - |

**Table S2.** Lattice constants for all compositions, found by X-ray diffraction measurements.

## Section S3. Magnetic properties

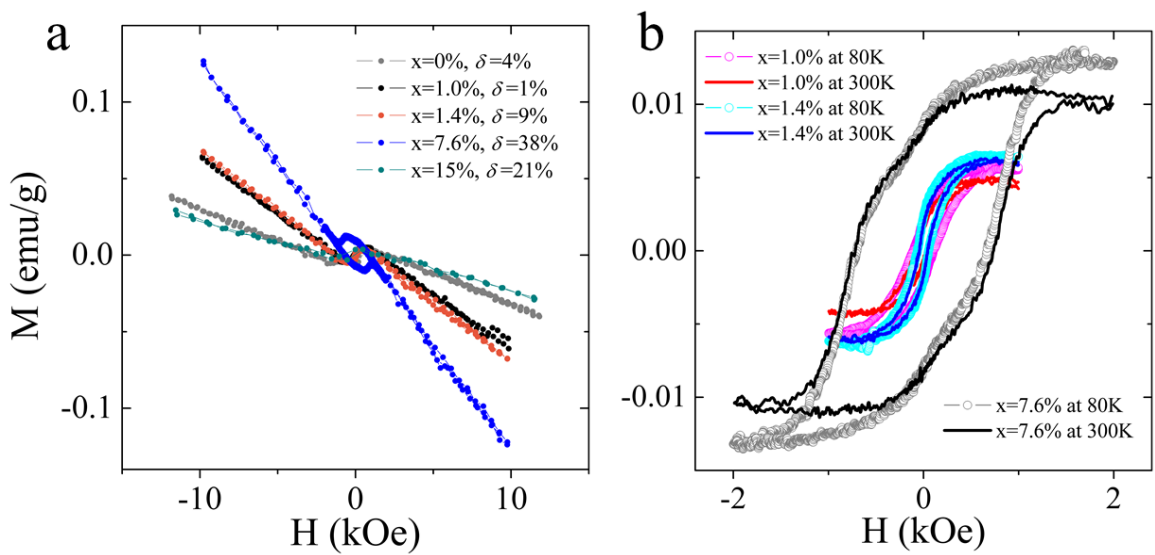

**Figure S3. (a)** Total magnetization measurements at 300 K for all tellurium-doped samples **(b)** Ferromagnetic contribution in the magnetization for tellurium-doped (x=1.0-7.6%) single crystals at 300K and 80K.

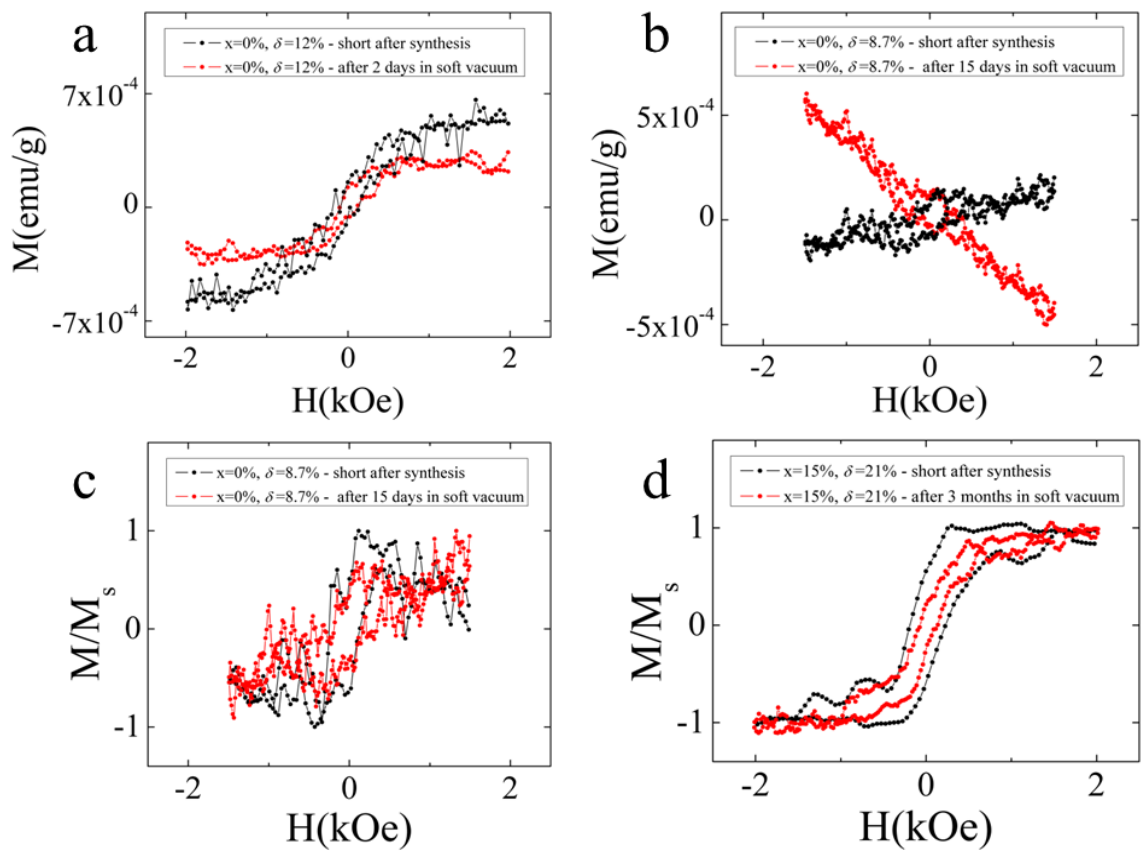

**Figure S4. (a,b)** Total magnetization hysteresis loops, for two different times of exposure to ambient conditions, for samples without Te-doping and vacancy level of **(a)** δ=12%, and **(b)** δ=8.7%. **(c,d)** Ferromagnetic contribution to the magnetization for two different times of exposure to ambient conditions, for **(c)** the undoped sample with δ=8.7% and **(d)** the Te doped sample with x=15%, δ=21%.

Fig. S3(a) shows the total magnetization in emu/g at 300K for x=0-15% tellurium-doped compositions. Here, the magnetic response reveals two components: a diamagnetic and a ferromagnetic contribution. The ferromagnetic component, obtained by subtracting the linear-diamagnetic contribution of each curve, has a slight temperature dependence, particularly in the saturation magnetization, as evidenced in fig. S3(b) which shows measurements at 80 K and 300 K for different Te-compositions.

On the other hand, the diamagnetic component has a strong dependence with the level of oxidation of the samples. Figs. S4(a,b) show the total magnetization curves for two types of samples, comparing the results for two different levels of exposure to ambient conditions: short after synthesis (short exposure-time to ambient conditions), and after several days of storage in soft vacuum. For the curves it is clear that the diamagnetic slope is developed only after a prolonged exposure to ambient conditions, whereas the ferromagnetic contribution of the magnetization is present with and without exposure to oxygen. Moreover, the ferromagnetic contribution seems to be more robust to slight oxidation of the samples, as shown in figs. S4(c,d). These curves reveal that the coercive field of the ferromagnetic hysteresis loops present only slight variations with the level of exposure to ambient conditions. Table S3 shows the variation in coercive field and saturation magnetization of the ferromagnetic contribution to magnetization, for different times of exposure to oxygen, for four different types of samples. Although the saturation magnetization presents variations with the time of exposure to oxygen, the coercive field is very robust to the oxidation level, and this is the reason why this quantity was the one chosen to study the evolution of the magnetism with Te-doping and chalcogen vacancy level.

| Magnetic property | | XFR Compositions | | | |
|---|---|---|---|---|---|
| | | x=0%<br>δ=12% | x=1%<br>δ=1% | x=0%<br>δ=8.7% | x=15%<br>δ=21% |
| | Storage time (days) | 2 | 3 | 15 | 90 |
| $H_c$ | Short after synthesis (Oe) | $113 \pm 25$ | $47 \pm 34$ | $199 \pm 20$ | $188 \pm 25$ |
| | After soft vacuum storage (Oe) | $113 \pm 10$ | $47 \pm 15$ | $169 \pm 40$ | $91 \pm 18$ |
| | (%) var | $0.00 \pm 0.04$ | $0.0 \pm 0.5$ | $-15 \pm 4$ | $-52 \pm 10$ |
| $M_s$ | Short after synthesis (memu/g) | $0.53 \pm 0.06$ | $8 \pm 1$ | $0.13 \pm 0.05$ | $2.7 \pm 0.6$ |
| | After soft vacuum storage (memu/g) | $0.28 \pm 0.07$ | $4.8 \pm 0.4$ | $0.12 \pm 0.06$ | $0.52 \pm 0.07$ |
| | (%) var | $-47 \pm 12$ | $-40 \pm 3$ | $-8 \pm 3$ | $-81 \pm 11$ |

**Table S3.** Variation in ferromagnetic properties after soft vacuum storage.

## Section S4. PFM measurements in periodically poled lithium niobate (PPLN)

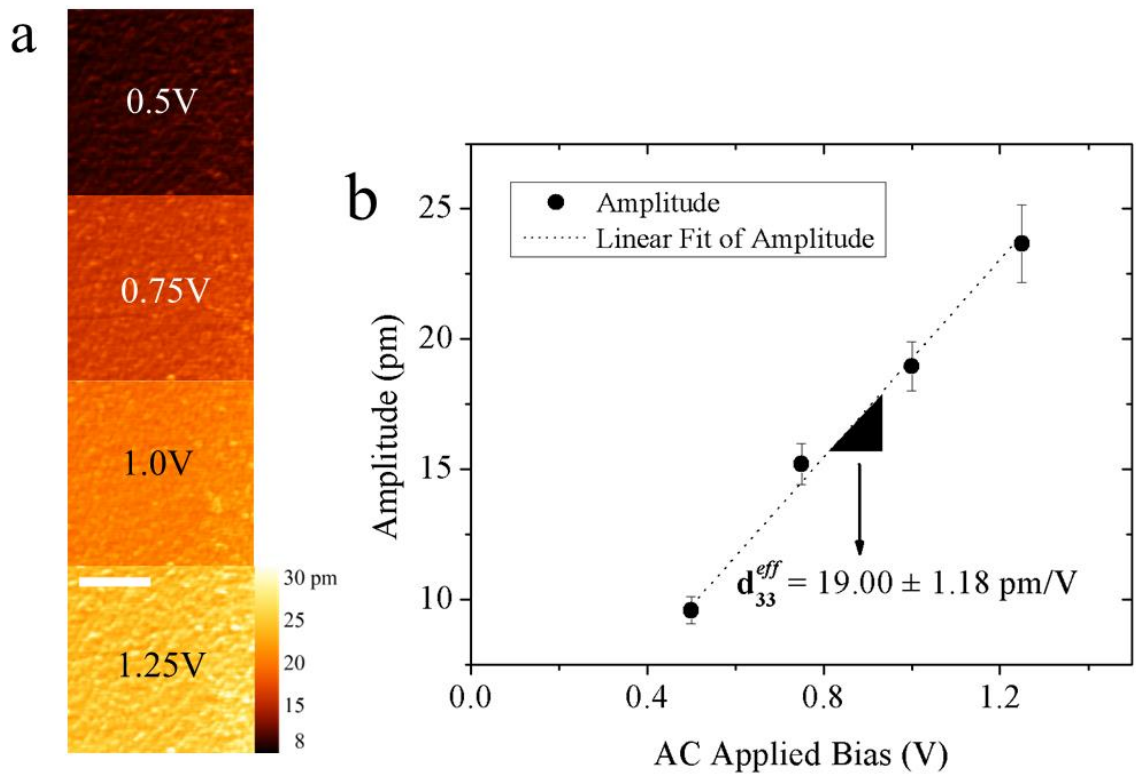

**Figure S5. (a)** DART-PFM piezoresponse normalized by SHO model as a function of the AC applied bias for PPLN material, and **(b)** their resultant effective $d_{33}$. Scale bar: 1 μm.

In order to check the calibration of our PFM experimental setup, and to gain confidence on the accuracy of the reported values for the piezoelectric coefficients of our single crystals, we performed DART-PFM and SS-PFM measurements in a commercially available Periodically Poled Lithium Niobate (PPLN) sample. Fig. S5(a) shows the results of the DART-PFM measurements, in which the deformation amplitude is scanned in an area of about 3 um x 3 um, and for different applied AC bias voltages, ranging from 0.5V to 1.25V. The amplitude values were normalized by the SHO model. The spatial average of deformation amplitude is plotted as a function of bias voltage in fig. S5(b). The slope of the linear regression to this curve corresponds to the piezoelectric coefficient, $d_{33}$. For our sample of PPLN, the experimental result $d_{33}$ is 19.00 ± 1.18 pm/V. This value is similar to the reported values for $d_{33}$ in multiple piezoelectric studies for PPLN (*27*). Therefore, this demonstrate that the presented method is reliable for the determination of effective piezoelectric constants.

Fig. S6 shows additional PFM measurements on the same PPLN sample. DART-PFM measurements (figs. S6(a-c)) allow the identification of poled domains in this material, which are not visible on the topography image. These domains were investigated by SS-PFM over the marked points in fig. S6(c). Firstly, both, +*z* poled and -*z* poled regions were measured using DC pulses bias from 0 to ±1V. The SS-PFM phase shows that these domains are oriented in opposite directions. Their SS-PFM amplitudes reveal an opposite behavior on the sample deformation: a contraction for -*z* poled and an extension for +*z* poled (fig. S6(d)).

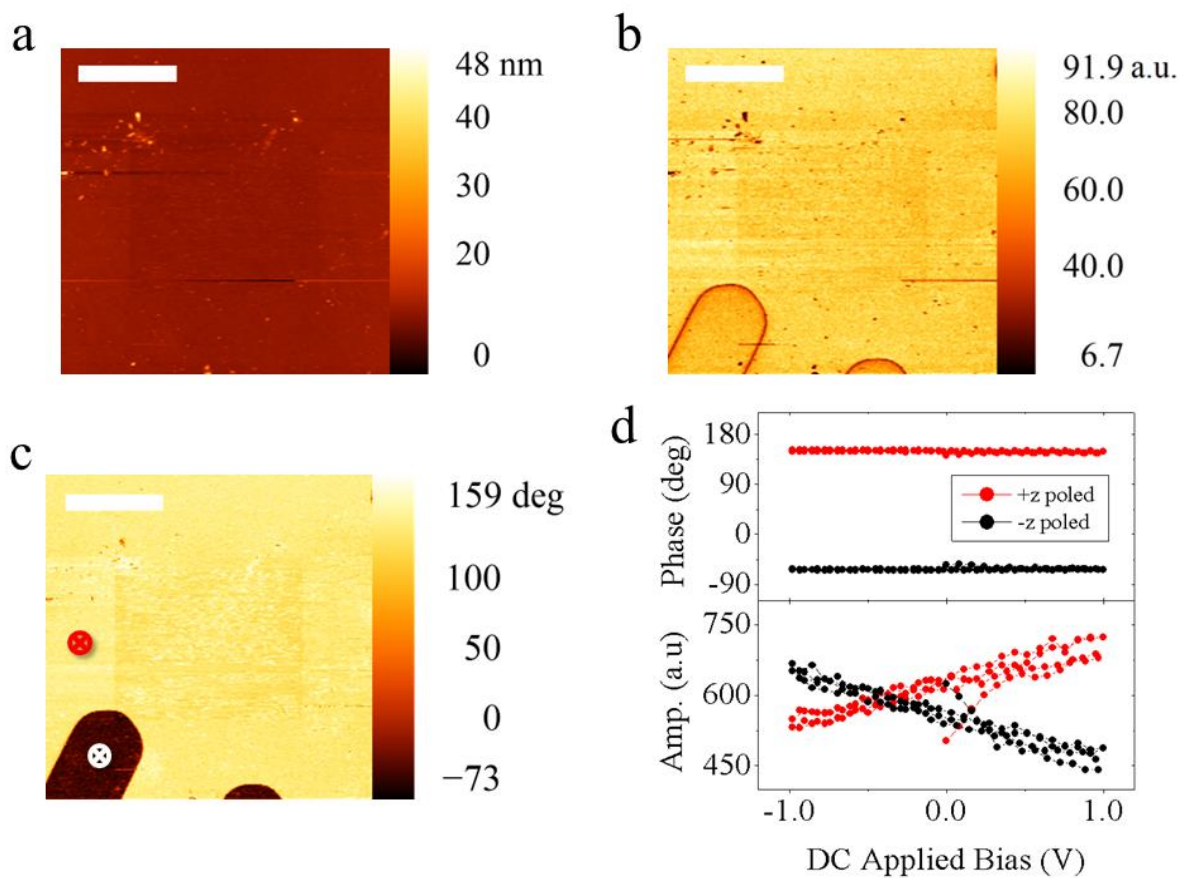

**Figure S6.** PFM measurements over Periodically Poled Lithium Niobate (PPLN) sample. **(a)** Topography and DART-PFM **(b)** amplitude and **(c)** phase. Scale bar: 5 μm. **(d)** SS-PFM amplitude and phase for a DC pulses bias from 0 to ±1V in ±$z$-poled points shown in (c).

## Section S5. Piezoresponse force microscopy hysteresis loops

Switching Spectroscopy (SS) PFM hysteresis loops for all tellurium-doped compositions, and an undoped sample (x=0%, δ=4%), are compared in fig. S7. Given that the undoped sample - which is not piezoelectric - was measured in the same conditions as the Te-doped samples, it can be used to identify the contribution of electrostatic effects during the measurement. It is worth noting that those contributions are negligible when compared to the experimental signal in our Te-doped compositions. Significantly, the amplitude of the SS-PFM signal increases enormously with increasing tellurium doping, which confirms the large enhancement of the piezoelectric coefficient for the highest Te-doping sample.

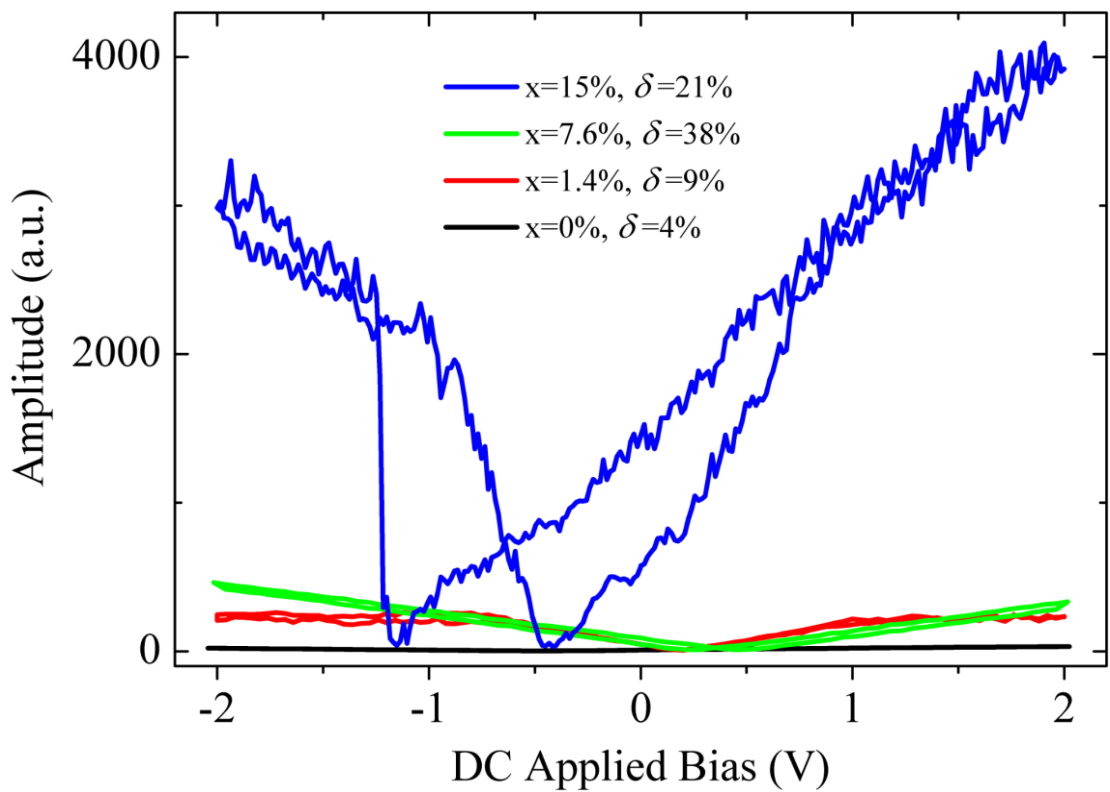

**Figure S7.** Switching Spectroscopy (SS) PFM hysteresis loops at 300K under an inert atmosphere for all tellurium-doped compositions. An undoped composition (black) is measured to illustrate contributions to the amplitude caused by electrostatic effects.

## Section S6. Transport properties

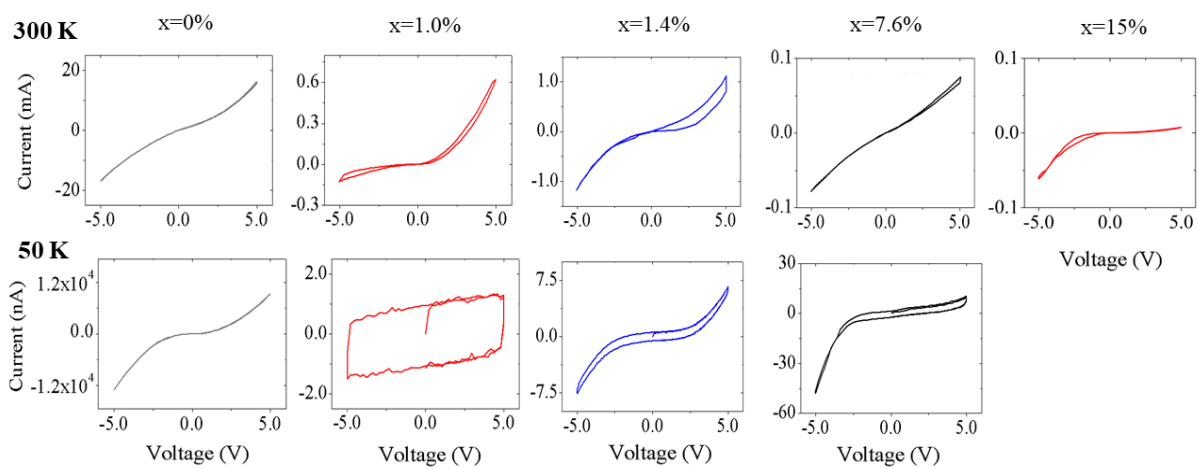

**Figure S8.** IV characteristics of undoped and Te-doped single crystals at 300K (upper row) and 50K (lower row).

Another indication of ferroelectric properties in our single crystals can be found in the current vs voltage (IV) curves. Fig. S8 shows I-V measurements taken in samples from x=0% to 15%, at 300K and 50K.

Undoped WSe$_{2(1-\delta)}$ samples show an S-shape symmetrical curve, characteristic of a semiconductor, both at 300K and 50K, with no signs of anomalies. In contrast, all other tellurium-doped samples show IV curves that are highly asymmetrical with respect to the voltage polarity (diode-like behavior), and with clear hysteretic behavior (resistive switching-like behavior, and/or capacitive hysteresis-like behavior). This type of behavior seems to be enhanced at low temperature for all Te-doped samples. It is worth highlighting that the hysteretic and asymmetric behavior present at 300K for all Te-doped samples are commonly observed in prototypical ferroelectric materials such as BaFeO$_3$ and BaTiO$_3$, and the origin of this multifunctional (i.e., diode-like, resistive switching and capacitive hysteresis effects) is an active area of research, with multifunctional applications in electronic devices.

## Section S7. DFT computed magnetization

From the table S4 it is clear that although SOC effects affect the magnetization, they are small and could in principle be ignored. The largest magnetization is obtained for the highest vacancy level, independently of the Te-doping.

| $x$ | $\delta$ | $a$(Å) | $c$(Å) | $M$ ($\mu_B$/ unit cell) - no SOC | $M$ ($\mu_B$/ unit cell) - SOC |
|---|---|---|---|---|---|
| 0.0000 | 0.0000 | 3.289 | 12.999 | 0.0000000 | 0.0000000 |
| 0.0000 | 0.0000 | 3.29(58) | 13.09(58) | - | - |
| 0.0000 | 0.31250 | 3.267 | 12.946 | 0.000000000 | 0.000000156 |
| 0.0000 | 0.109375 | 3.209 | 12.813 | 0.000000000 | -0.000000625 |
| 0.0000 | 0.390625 | 2.964 | 12.009 | 0.189400000 | 0.190000000 |
| 0.015625 | 0.0000 | 3.293 | 13.044 | 0.000000000 | 0.000000000 |
| 0.0625 | 0.0000 | 3.303 | 13.118 | 0.000000000 | 0.000000000 |
| 0.017544 | 0.109375 | 3.212 | 13.184 | 0.000000000 | 0.000000000 |
| 0.102564 | 0.390625 | 2.961 | 13.092 | 0.177000000 | 0.177000000 |

**Table S4.** Computed total unit cell magnetization, for different levels of Te-doping and chalcogen vacancies, with and without SOC.

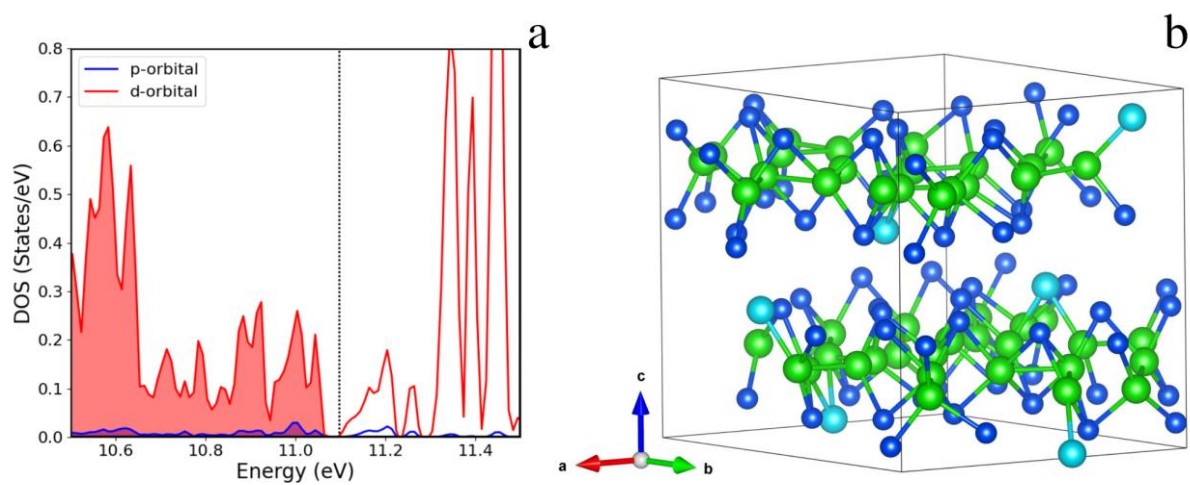

**Figure S9. (a)** Projected density of states on atomic orbitals of W1, the atom with the largest magnetic moment. **(b)** Optimized unit cell for a second configuration of W(Se$_{1-x}$Te$_x$)$_{2(1-\delta)}$ with $x$=10.26% and $\delta$=39.06%. In this case all W atoms are bonded to at least one chalcogen atom and the resulting magnetization is zero.

## Section S8. DFT computed electric polarization

| $x$ | $\delta$ | Hirshfeld charge ($10^{-2}$ a.u.) | $P_x$ ($\mu C/cm^2$) | $P_y$ ($\mu C/cm^2$) | $P_z$ ($\mu C/cm^2$) |
|---|---|---|---|---|---|
| 0.0000 | 0.0000 | $Q_W$=-1.57<br>$Q_{Se}$= 0.79 | 0.00 | 0.00 | 0.00 |
| 0.102564 | 0.0000 | | 4.09E-08 | -1.51E-07 | -7.98E-07 |
| 0.0000 | 0.390625 | $Q_{W27}$=-7.46<br>$Q_{W28}$=-3.26<br>$Q_{W32}$=-10.33<br>$Q_{Se7}$=5.54 | 9.80 | 0.05 | -9.95 |
| 0.102564<br>Se5<br>below<br>vacancy | 0.390625 | $Q_{W27}$=-5.30<br>$Q_{W28}$=-2.52<br>$Q_{W32}$=-9.37<br>$Q_{Se5}$=3.31 | -5.43 | -4.06 | -0.36 |
| 0.102564<br>Se5<br>above<br>vacancy | 0.390625 | $Q_{W27}$=-5.86<br>$Q_{W28}$=-2.80<br>$Q_{W32}$=-9.62<br>$Q_{Se5}$=4.49 | 7.16 | -0.44 | 6.04 |
| 0.102564<br>Te2<br>below<br>vacancy | 0.390625 | $Q_{W27}$=-4.82<br>$Q_{W28}$=-4.20<br>$Q_{W32}$=-10.18<br>$Q_{Te2}$=8.64 | 0.58 | -8.65 | -5.54 |
| 0.16 | 0.2187 | | -0.71 | 3.12 | 4.7 |

**Table S5.** Computed values for Hirshfeld charge of different atoms in the unit cell, and total unit cell electric polarization, for different values of Te-doping and chalcogen vacancies.

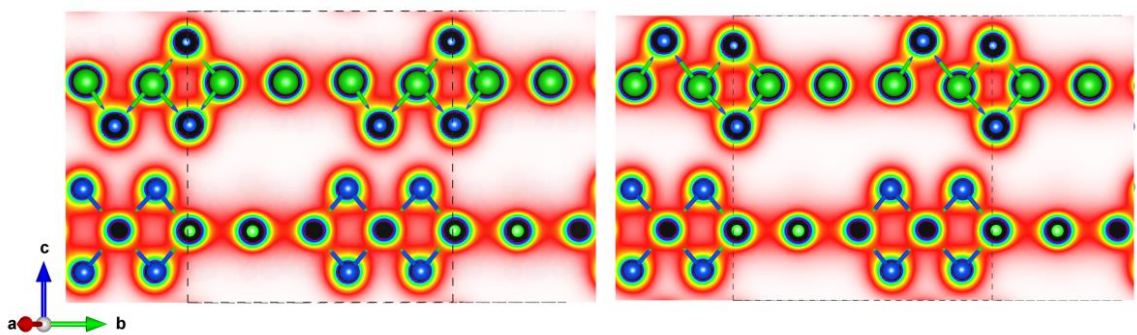

**Figure S10.** [1 0 0] views of the total electronic densities as obtained for the vacancy on top (left) and below (right) a Se atom.

## Section S9. List of suporting files.

The crystallographic information filed - CIF - of the resolved crystal structure found by single crystal diffraction, for two of the studied compositions, can be found in the files:

| Te-concentration | Files |
|---|---|
| $x$=1% | File S1: S1_WSeTe-1percent.cif<br>File S2: S2_checkcif report for WSeTe-1percent.pdf |
| $x$=7.6% | File S3: S3_WSeTe-7percent.cif<br>File S4: S4_checkcif report for Te-7percent.pdf |

Computed atomic coordinates, lattice constants as well as Hirshfeld charge analysis for several configurations can be found in the following files:

| Configuration | Files |
|---|---|
| $x$=6.25%, $\delta$=39.0625% and Se5 atom below the vacancy. | File S5: S5_WSe2Te6.25Vac39-relaxed.xsf<br>File S6: S6_Hirshfeld-Te6.525Vac39.txt |
| $x$=6.25, $\delta$=39.0625% and Se5 atom above the vacancy. | File S7: S7_WSe2Te6.25Vac39-SeAbove.xsf<br>File S8: S8_Hirshfeld-Te6.525Vac39-SeAbove.txt |
| $x$=6.25%, $\delta$=39.0625% and Te2 atom below the vacancy. | File S9: S9_WSe2Te6.25Vac39-TeBelow.xsf<br>File S10: S10_Hirshfeld-Te6.25Vac39-TeBelow.txt |
| $x$=6.25%, $\delta$=39.0625% non-magnetic configuration. | File S11: S11_WSeTeVac-conf2-relaxed.xsf |
| $x$=0%, $\delta$=39.0625% and Se5 atom below the vacancy. | File S12: S12_WSe2Te0.0Vac39.xsf<br>File S13: S13_Hirshfeld-Te0.0Vac39.txt |
| $x$=16%, $\delta$=21.87% | File S14: S14_WSeTe12.5Vac21.87-relaxed.xsf<br>File S15: S15_Hirshfeld-WSeTe12.5Vac21.87.txt |